\documentclass[superscriptaddress,prl,twocolumn,showpacs]{revtex4}

\usepackage{graphicx}
\usepackage{dcolumn}
\usepackage{epsfig}
\usepackage{latexsym}
\usepackage{amsfonts}
\usepackage{amssymb}

\begin{document}
\bibliographystyle{apsrev}
\title{Entropy driven formation of a chiral liquid crystalline phase of helical filaments}

\author{Edward Barry}\affiliation{Rowland Institute at Harvard, Harvard University,
Cambridge, MA 02142 }

\author{Zach Hensel}\affiliation{Rowland Institute at Harvard, Harvard University,
Cambridge, MA 02142 }
\author{Michael Shribak}\affiliation{Marine Biological Laboratory, Woods Hole, MA 02543}

\author{Rudolf Oldenbourg}\affiliation{Marine Biological Laboratory, Woods Hole, MA 02543}

\author{Zvonimir Dogic} \affiliation{Rowland Institute at Harvard, Harvard University,
Cambridge, MA 02142 }

\date{\today}
\begin{abstract}
We study the liquid crystalline phase behavior of a concentrated
suspension of helical flagella isolated from {\it Salmonella
typhimurium}. Flagella are prepared with different polymorphic
states, some of which have a pronounced helical character while
others assume a rod-like shape. We show that the static phase
behavior and dynamics of chiral helices are very different when
compared to simpler achiral hard rods. With increasing
concentration, helical flagella undergo an entropy driven first
order phase transition to a liquid crystalline state having a
novel chiral symmetry.
\end{abstract}


\maketitle

Molecules with chiral symmetry cannot be superimposed on their
mirror image. Such molecules can assemble into a variety of
complex chiral structures of importance to both physics and
biology~\cite{Harris99}. Chirality has a special prominence in the
field of liquid crystals and the presence of a chiral center can
dramatically alter the liquid crystalline phase behavior and its
material properties~\cite{Renn88}. For example, achiral rods form
a nematic phase with long range orientational order. However,
rearranging a few atoms to create a microscopically chiral
molecule can transform a nematic into a cholesteric phase.
Locally, a cholesteric phase has a structure in which molecules
are organized in layers. Within a layer the molecules are parallel
to each other, while the molecular orientation is slightly rotated
between two  adjacent layers. This order at least partially
satisfies the pair interaction between neighboring chiral
molecules which tends to twist their mutual orientation. Even for
the fairly simple example of a cholesteric phase, it is difficult
to establish a rigorous relation between the microscopic chirality
of the constituent molecules and the macroscopic chirality
characterized by the cholesteric pitch~\cite{Harris97,Straley76}.

In stark contrast to our poor understanding of the cholesteric
phase, microscopic theories of nematic liquid crystals have been
very fruitful~\cite{Onsager49}. Onsager realized that a simple
fluid of concentrated hard rods will form a stable nematic phase.
Using his theory it is possible to predict the macroscopic phase
behavior of a nematic suspension of hard rods from microscopic
parameters such as rod concentration and rod aspect ratio. Due to
the dominance of repulsive interactions, phase transitions within
the Onsager model belong to a class of entropy driven phase
transitions. Inspired by the success of the Onsager theory,
Straley made the first attempt at formulating a microscopic theory
of the cholesteric phase~\cite{Straley76}. In this work hard-rod
interactions are extended to threaded rods which have an
appearance similar to screws. The excluded volume between two
threaded rods is at a minimum not when they are parallel to each
other but when they approach each other at an angle at which their
chiral grooves can interpenetrate. This results in the formation
of a cholesteric phase that is, as in the Onsager model, entirely
driven by entropic excluded volume interactions.

Biopolymers such as DNA, actin, TMV and {\it fd} are good
experimental system to study certain aspects of liquid crystalline
ordering of colloidal rods~\cite{Livolant91,Podgornik96,Grelet03}.
In particular, monodisperse rod-like viruses were used to
quantitatively test the Onsager
theory~\cite{Grelet03,Purdy03,Oldenbourg88}. However, the
usefulness of these systems in elucidating the microscopic origin
of chirality is severely limited due to the inability to precisely
control their microscopic chiral structure. While almost all
biopolymers have chiral stucture, some of them form a nematic
phase ({\it pf1} and TMV), while others form a cholesteric phase
({\it fd}, PEG-{\it fd} and DNA)~\cite{Livolant91,Grelet03}. The
reason for this behavior remains a mystery.

\begin{figure}
\centerline{\epsfig{file=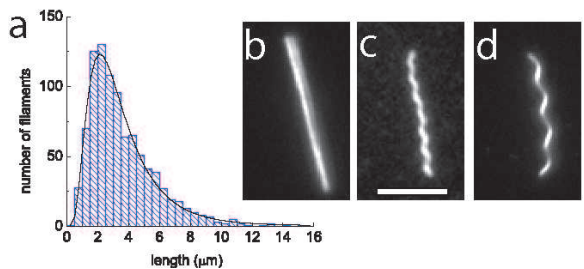,width=3in}}
\caption{\label{figure1} (a) Length distribution of flagella
isolated from strain SJW1103. (b) Fluorescently labelled straight
flagella isolated from strain SJW1665. (c) Mutant flagella
isolated from SJW2869. Helical flagella are characterized by their
pitch (P=1.1 $\mu$m) and diameter (D=0.160 $\mu$m). The average
length is 4.01$\pm$2.5 $\mu$m. (d) Wild type flagella isolated
from strain SJW1103 (P=2.4 $\mu$m, D=0.4 $\mu$m). The average
length is 3.6 $\pm$2.2 $\mu$m. Scale bar indicates 5 $\mu$m. }
\end{figure}

In this letter we use flagella isolated from prokaryotic bacteria
to experimentally study the phase diagram of a concentrated
suspension of helices. Flagellar filaments are macromolecular
structures assembled from a single protein called flagellin. A
unique advantage of a bacterial flagellum is that its helical
shape is precisely regulated and depends on the flagellin amino
acid sequence and the pH and temperature of the suspension
(Fig.~\ref{figure1})~\cite{Kamiya80}. Moreover, it can be tuned in
a number of discrete steps from essentially achiral hard rods to
highly twisted helices. Using this system we can readily control
the microscopic chirality of the flagella and relate this to its
liquid crystalline phase behavior. Unlike the predictions of the
Straley model, we find that the liquid crystalline behavior of
helices is dramatically different from that of straight rods. They
form a liquid crystalline phase with a novel chiral symmetry which
we call a conical phase. The symmetry of this phase is similar to
a structure that can occur when a cholesteric phase is placed in a
magnetic field along its twist direction and the molecules
consequently tip along the applied field~\cite{Meyer68}.
Experimentally this case is rarely observed since Frank elastic
constants have to satisfy the condition $K_{22}>K_{33}$. We also
show that the dynamics of helices in a conical phase is very
different from the dynamics of achiral rods in a nematic. Previous
studies of ordering in flagella was limited to achiral rods for
the purpose of structure determination~\cite{Yamashita98}.

The phase diagram of flagella suspensions is highly dependent on
their length distribution, which in turn is sensitive to
purification methods. The flagellar filaments are isolated from
{\it S. typhimurium} strains SJW 1103, SJW 2869 and SJW
1664~\cite{Kanto91}. Bacteria grown to a log-phase is sedimented
at 8000g and redispersed by repeated pipetting with 1 mL pipette.
Subsequently, a uniformly foamy solution is vortexed at the
highest power setting for 5 minutes (Genie 2 Vortex, VWR).
Vortexing shears long flagellar filaments from cell bodies. The
filaments are separated from much heavier cells by a low speed
centrifugation step at 8000g for 20 minutes. The supernatant
contains flagellar filaments, which are purified and concentrated
by two centrifugation steps at 100,000g for 1 hour. This process
yields about 50 mg of flagelllar filaments. Filaments are labelled
with a fluorescent dye (TAMRA-SE, Molecular Probes). To obtain the
histogram of filament lengths, we dry a suspension of flagella in
1\% methyl-cellulose onto a coverslip which flattens 3D helices.
Unless otherwise noted all the images were taken with a Nikon
polarization microscope E600 equipped with fluorescence
components.

\begin{figure}
\centerline{\epsfig{file=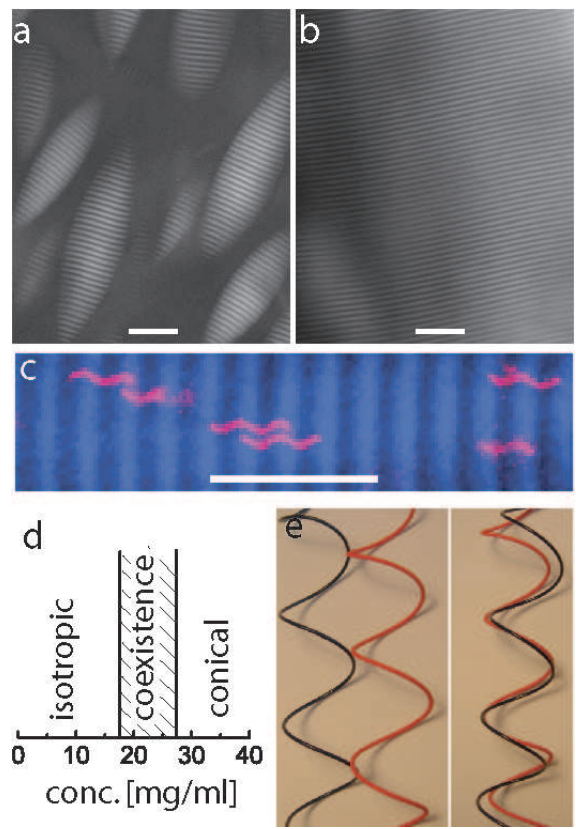,width=7.5cm}}
\caption{\label{figure2} (a) Coexistence between isotropic and
conical phase in flagella SJW1103 at 20 mg/ml imaged with
polarization microscope. Scale bar is 20 $\mu$m.  (b) A single
phase imaged with polarization microscope, scale bar is 20 $\mu$m.
(c) Fluorescently labelled flagella dissolved in a conical phase
of unlabelled flagella. Fluorescent image was overlayed over a
polarization microscope image. Scale bar is 10 $\mu$m. (d) Phase
diagram of flagella SJW1103 as a function of flagella
concentration. (e) Illustration of excluded volume between two
helical rods out of phase and in phase with respect to each
other.}
\end{figure}

At concentrations below 17 mg/ml, wild-type flagella SJW1103 form
isotropic suspensions. Above this concentration, samples initially
appear in a uniformly birefringent metastable state. Over a period
of days bright birefringent droplets (tactoids) start phase
separating from a dark isotropic background (Fig.~\ref{figure2}a).
These droplets exhibit a well defined striped pattern with a 2.4
$\mu$m periodicity. With increasing flagella concentration, the
volume fraction of tactoids increases and at concentrations above
29 mg/ml the samples appear birefringent everywhere with no hint
of coexisting isotropic solution. These single phase samples have
a polydomain texture with a very small domain size. Moreover, the
domains do not anneal over time. The best way to obtain
macroscopically aligned samples (Fig.~\ref{figure2}b) is to
prepare a coexisting sample and let it bulk phase separate over a
period of a few weeks. From these observations we obtain a
tentative phase diagram of flagella suspensions shown in
Fig.~\ref{figure2}d. The slow dynamics and possible presence of a
non-equilibrium glassy phase complicate the accurate determination
of the uppper binodal point. The large width of the
isotropic-liquid crystalline coexistence is a consequence of the
polydispersity of the filaments~\cite{Wensink03}. Over a
temperature range from 0 to 40 $^\circ$C the phase diagram does
not change, indicating that temperature independent repulsive
interactions dominate the phase behavior. The well defined shape
of the tactoids indicate that these are equilibrium structures and
are not formed due to irreversible aggregation. Tactoids such as
these are frequently observed in coexisting samples of a wide
variety of rod-like colloids~\cite{Prinsen03}.

\begin{figure}
\centerline{\epsfig{file=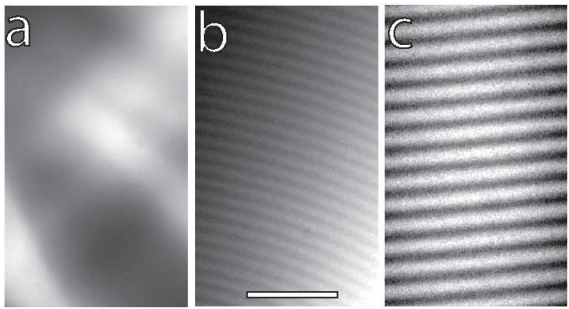,width=7.5cm}}
\caption{\label{figure3} Polarized light images of liquid
crystalline flagella suspensions. (a) Nematic liquid crystal phase
observed in straight flagella SJW1665. (b) and (c) Conical phase
formed from helical flagella SJW2869 and SJW1103 respectively.
Bright and dark stripes indicate differing director orientations.
Images in b and c exhibit one dimensional periodicities of 1.1 and
2.4 $\mu$m respectively. Scale bar is 10 $\mu$m.}
\end{figure}

The most surprising feature of the liquid crystalline phase of
flagella SJW1103 is a well defined one-dimensional periodicity
with a wavelength that corresponds to the pitch of its constituent
flagella. The striped pattern is only observed with polarization
microscopy which is sensitive to the local orientation of the
molecules. When viewed with DIC microscopy, which is sensitive to
concentration differences, no spatial variations are observed.
Therefore there are only orientational variations and no density
variations along the long axis of the tactoid. In addition,
fluorescent images reveal that the filaments are always in phase
with respect to each other (Fig.~\ref{figure2}c). These
observations lead us to propose a liquid crystalline structure in
which the helical filaments are intercalated with each other and
the liquid crystalline director follows the path set by the
geometry of constituent molecules. While there are no 1D density
variations associated with this order, there are clear one
dimensional orientational variations which make the director span
the surface of a cone. Therefore, we call this liquid crystalline
phase a conical phase which was first described in
Ref.~\cite{Meyer68}. Polydispersity of the helices suppresses the
formation of a positional smectic. We emphasize that this
organization of molecules is very different from cholesterics
where the director twists perpendicularly to its local direction.
It is intuitively clear how excluded volume interactions result in
the formation of a conical phase. The excluded volume between two
helices is much larger when they are out of phase then when they
are in phase with each other (Fig.~\ref{figure2}e). This is the
driving force for the formation of the conical phase.

Figure~\ref{figure3} illustrates the behavior of the liquid
crystalline phase as the geometry of the filaments changes. As
expected, straight filaments form a nematic (Fig.~\ref{figure3}a).
Slightly helical filaments (strain SJW2869) form a conical phase
with spacing of 1.1 $\mu$m, corresponding to the pitch of an
isolated flagelum. For this sample as well as for flagella
SJW1103, we observe a direct transition from isotropic to conical
phase with no intermediate nematic phase. It is also possible to
envision a scenario in which helical rods first form a nematic
(cholesteric) phase and only upon further compression, when
helices begin to intercalate, undergo a second transition to a
conical phase. The flagella SJW2869 exhibit an isotropic phase
below 7 mg/ml, coexisting samples between 7 mg/ml and 26 mg/ml and
a conical phase above 26 mg/ml. The liquid crystalline phase in
SJW2869 occurs at much lower concentrations when compared to
SJW1103 despite similar length distribution. This might be due to
the larger ratio of contour length over diameter of flagella
SJW2869.

\begin{figure}
\centerline{\epsfig{file=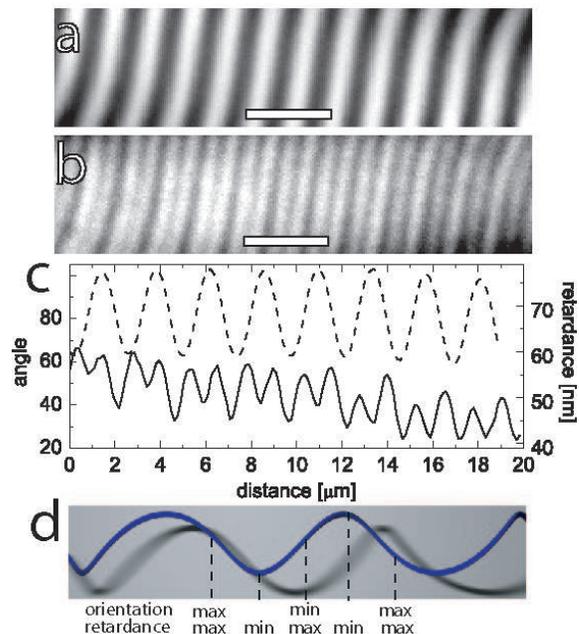,width=7.5cm}}
\caption{\label{figure4} Analysis of a conical phase of flagella
SJW1103 based on LC-PolScope images. (a) and (b) Images show a map
of the orientation of the slow axis of the birefringence and a
retardance map, respectively. Scale bars are 5 $\mu$m. (c) Line
profile along the striped pattern of the orientation image (dashed
line) and retardance image (full line). When comparing the
retardance to orientation image frequency doubling is observed.
(d) Dashed lines represent the location along a helical director
where the minimum and maximum of the orientation and retardance
occurs. From this schematic it is clear that retardance has twice
the periodicity of orientation.}
\end{figure}

To gain a better understanding of the conical phase we have
analyzed our samples with quantitative 2D polarization microscopy
(LC-PolScope) which simultaneously provides detailed spatial maps
of the retardance and orientation of the slow axis of
birefringence~\cite{Oldenbourg95}. Similar to traditional
polarized light images, the retardance and orientation maps show
striped patterns indicating the systematic variations in director
orientation. In some cases, the striped pattern visible in the
retardance image (Fig.~\ref{figure4}b) exhibits half the period
(or twice the spatial frequency) compared to the orientation image
(Fig.~\ref{figure4}a). This is consistent with the liquid
crystalline director following a helical trajectory in the image
plane. By way of explanation, the measured retardance in an
LC-PolScope image is modulated by the inclination of the director
with respect to the object plane that is in focus. When the
director is inclined to the object plane, light travels more
parallel to the director, and the measured retardance decreases
compared to when the director is parallel to the object plane.
When the director follows a helical path, with the helix axis
parallel to the object plane, the director attains a parallel
orientation twice within one helical pitch and attains positive
and negative inclination angles in-between. Because the reduction
is the same for positive and negative inclination angles, the
observed periodicity in the retardance image is half the helical
pitch. If the helix axis is not parallel to the object plane, the
magnitude of the extreme positive and negative inclination angles
are unequal and the retardance minima are unequal, as observed in
Figure 4 b and c. It is interesting to note that the appearance of
twice the fundamental frequency is usually not apparent in images
recorded using a traditional polarization microscope (Figs. 2 and
3). The LC-PolScope generates independent retardance and
orientation maps and makes the doubling of the frequency in
retardance images visible.

\begin{figure}
\centerline{\epsfig{file=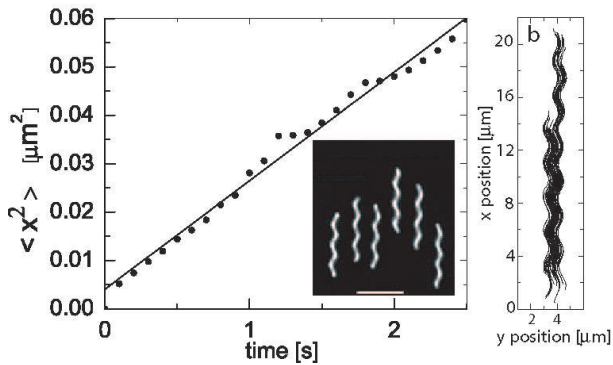,width=8.0cm}}
\caption{\label{figure5} Dynamics of a fluorescently labelled
flagellum in a conical phase. The flagellum exhibits effective
translation motion through rotation similar to the behavior of a
``nut on a bolt.''  (a) Mean square displacement of the phase
angle of an individual flagellum. Inset: A time sequence of a
diffusing flagellum taken over a period of several minutes. (b)
Overlapped trajectories reconstituted of a flagellum diffusing in
a conical phase.}
\end{figure}

In nematics the broken orientational symmetry results in
anisotropic diffusion coefficients with rods preferentially
diffusing along the nematic director~\cite{Lettinga05}. In a
similar fashion, the helical symmetry of the conical phase results
in an unusual diffusion of constituent helices. Fig.~\ref{figure5}
shows a time sequence and reconstructed trajectories of an
isolated flagellum diffusing in a conical phase. As shown in
Fig.~\ref{figure5}b all the peaks and valleys are in register,
indicating that translational diffusion is strictly coupled to
rotational diffusion. This is similar to the movement of a ``nut
on a bolt''. For diffusion along the helix axis the mean square
displacements of the filament endpoints increase linearly with
time leading to a translational diffusion constant of 0.023
$\mu$m$^2$/sec. For a translational displacement that corresponds
to one helical pitch, the filament rotates by 2$\pi$. Hence, the
translational diffusion of 0.023 $\mu$m$^2$/sec corresponds to a
rotational diffusion of 0.16 rad$^2$/sec. We find that the
diffusion depends on the length of the filaments with smaller
filaments diffusing faster then longer ones. A more detail study
of these phenomena will be presented elsewhere. In addition to
diffusion along the helical axis, we observe significant lateral
diffusion, which excludes the presence of columnar-like long range
order.

In conclusion, we note that the behavior of hard rods, the
simplest model system of liquid crystals, has been studied
extensively several decades. In the present paper we use well
regulated helical filaments to demonstrate that the static phase
behavior and dynamics of a concentrated suspension of helical
filaments is remarkably different from those of a suspension of
hard rods. Our work addresses the fundamental question of the
molecular origins of the varied liquid crystalline mesophases that
are found in biological systems and are believed to play an
important role in their function.

We thank Eric Grelet, Linda Stern, Tom Lubensky, Seth Fraden and
Randy Kamien for useful comments. Bacterial strains were kindly
provided by Linda Stern and Noreen Francis. ZD is supported by a
Junior Fellowship from Rowland Institute at Harvard. MS and RO are
supported by NIH grant Nr. EB002583.

\end{document}